\documentclass[prb,jcp,reprint,longbibliography]{revtex4-1}

\usepackage{amsmath}  
\usepackage{amsfonts} 
\usepackage{graphicx} 

\begin{document}



\title{Log Complex Color for Visual Pattern Recognition of Total
Sound}


\author{Stephen Wedekind}
\email[]{wedekind.stephen@gmail.com}
\author{P. Fraundorf}
\affiliation{Physics \& Astronomy/Center for Nanoscience, U. Missouri-StL (63121) USA}


\date{\today}

\begin{abstract}

While traditional audio visualization methods depict amplitude intensities vs. time, such as in a time-frequency spectrogram, and while some may use complex phase information to augment the amplitude representation, such as in a reassigned spectrogram, the phase data are not generally represented in their own right. By plotting amplitude intensity as brightness/saturation and phase-cycles as hue-variations, our complex spectrogram method displays both amplitude and phase information simultaneously, making such images canonical visual representations of the source wave. As a result, the original sound may be precisely reconstructed (down to the original phases) from an image, simply by reversing our process. This allows humans to apply our highly developed visual pattern recognition skills to complete audio data in new way.  (Published after peer review in 2016 as {\em Audio Engineering Society Convention} {\bf 141} paper 9647; now the subject of US patent 10,341,795.)

\end{abstract}
\pacs{81.05.uf, 06.20.fa, 06.20.Jr, 61.48.Gh}
\maketitle

\tableofcontents

\section{Introduction}
\label{sec:Intro}

While some current audio visualization methods use the complex\cite{Fritz2009, Fulop2006, Xiao2007, Flandrin2010, Flandrin2003} fast Fourier transform (FFT) components to augment the accuracy of (real) amplitude readings, they tend to be highly application-specific, and do not appear concerned with the significance of generalized, total-sound analysis, by which simultaneous display of both amplitude and phase data in each pixel provides a canonical means of recording, analyzing, cataloguing, and displaying more sound than humans are generally considered capable of hearing. Our work has been to develop an efficient and robust real-time method of viewing total sound spectrographs that incorporates log-intensity (for improved dynamic range) amplitude-visualization combined with chroma-like\cite{Bartsch2001, Cho2010, Harte2005} phase-visualization. By simultaneously displaying both real and imaginary FFT data-sets, we ensure that an image contains all the information of the original source, which means it is always possible to recover the original sound from any image generated with this method, down to the original phases. This opens the possibility of alternative data storage techniques, novel cataloguing methods such as visual sound field-guides (which, when combined with a mobile real-time visualization app could allow for live imitation-feedback), improved sound-availability for the hearing-impaired, and more. Additional modifications that include, e.g., Grand Staff musical overlay and/or stereo versions for wearable devices could help music readers without specific technical backgrounds and/or sensory capabilities to make sense of such total-sound visualizations. The ever-increasing capability of modern mobile devices can already support implementation of this visualization method, leveraging their wide distribution\cite{Wu2015} as well as their pre-installed microphones, color displays, and processing speeds\cite{Ukidave2014}.

\section{Methods}
\label{sec:Methods}

Our experience studying spatial periodicities in nanocrystalline solids\cite{pf2004m, pf2007m, pf2006n, Cowtan2014} has shown us the utility of representing both amplitude and phase with a single pixel, since condensed matter crystals contain periodicities in two and three spatial dimensions, and so require higher dimensional FFTs rather than the one time-dimension periodicities involved in audio analysis. By applying this visualization method to audio signals, we can display the complete, complex FFT of a given time-slice as a single column of pixels, allowing the horizontal axis to remain available for sequential slices in the time domain.

In contrast to current audio visualization methods
like traditional spectrograms, reassigned
spectrograms\cite{Fritz2009, Fulop2006, Xiao2007, Flandrin2010, Flandrin2003}, constant-Q transforms (CQTs)\cite{Brown1991, Brown1992, Velasco2011, Schorkhuber2010}, and chroma features\cite{Bartsch2001, Cho2010, Harte2005} which use
various techniques to optimize amplitude
visualization, we explore here only a simpler scheme
based on complex FFTs that simultaneously displays
the amplitude and phase information associated with
each pixel. As in many other applications\cite{Pucik2014, Bank2011, Harma2001, Jin2012},
not least of which is the traditional, Western musical
notation\cite{Mason1960, Sundberg1982, Cazden1958}, we optionally adopt logarithmic
scaling of the frequency axis since on it octaves and
harmonics are equally spaced. While techniques like
reassigned spectrograms utilize the imaginary part of
the Fourier transform to enhance accuracy of
particular amplitude and harmonic representations\cite{Fritz2009, Fulop2006, Xiao2007, Flandrin2010, Flandrin2003}, and chroma visualizations show periodic
changes in tone as hue-variations\cite{Bartsch2001, Cho2010, Harte2005}, our
method simultaneously displays both real and
imaginary Fourier data to produce a canonical view
of total sound. By showing Fourier coefficient
amplitude as the brightness/saturation of the
associated pixel, and Fourier phase as hue, each
pixel simultaneously represents both real and
imaginary components of a complex Fourier
coefficient.

\begin{figure}
\includegraphics[scale=2.8]{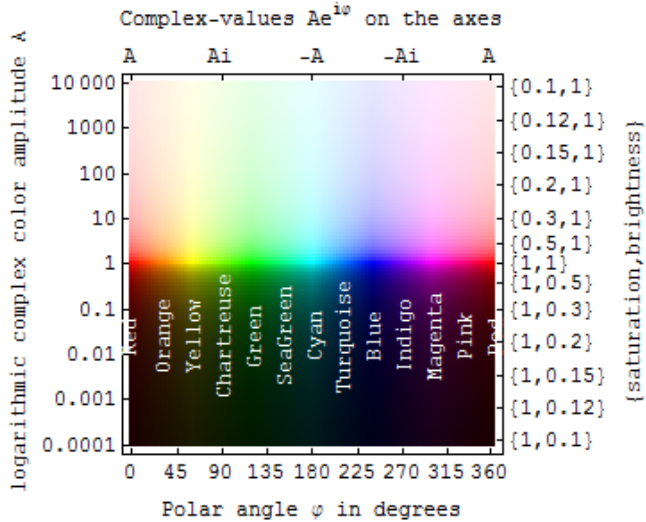}%
\caption{Logarithmic complex-color key in polar
coordinates with amplitude on the logarithmic
vertical axis and imaginary phase angle $\phi$ on the
linear horizontal axis.}
\label{fig1}
\end{figure}

On a linear frequency scale, log-color phase representation
begins with each complex Fourier
coefficient being converted to a color according to
Fig. \ref{fig1}. In such a representation, the hue is
determined by the coefficient's phase angle whereas
the brightness/saturation is determined by the
logarithm of the intensity of the coefficient. As seen
in Fig. \ref{fig1}, Fourier-coefficient phase-shifts in one
direction result in a red-to-green-to-blue (RGB)
sequence, whereas movement in the opposite
direction results in a red-to-blue-to-green (RBG)
sequence. Since the frequency scale is linear, the
only interpolation involved is that which maps the
saturation and brightness from a linear to a
logarithmic intensity scale (vertical axis of Fig. \ref{fig1}).
By plotting the log of the intensity rather than only
the intensity, some fine details are sacrificed in order
to provide conventional\cite{Giannoulis2013} improvements in
dynamic range. Hue, saturation, and brightness
parameters between 0 and 1 are determined by
equations (1), (2), and (3), respectively. This reversible mapping between complex-number
absolute-value and pixel-color thereby trades
contrast for dynamic range.

\begin{equation}
\text{hue} = \frac{\phi}{2 \pi}
\label{hue}
\end{equation}
\begin{equation}
\text{saturation} = 
 \begin{cases} 
  1 & \mbox{ if } A \le 1\\ 
   \frac{1}{1 + \ln[A]} & \mbox{ if } A > 1
 \end{cases}
\label{saturation}
\end{equation}
\begin{equation}
\text{brightness} = 
 \begin{cases} 
   \frac{1}{1 - \ln[A]} & \mbox{ if } A \le 1\\ 
   1 & \mbox{ if } A > 1
 \end{cases}
\label{brightness}
\end{equation}

In order to achieve the benefits of the log-frequency
scale from equally spaced samples in the timedomain,
the linear-frequency data must be
transformed, limiting the retention of some detailed
sound information in favor of a more robust visual
representation. In particular, since the transformation
from linear- to log-frequency expands the lowerfrequency
coefficients and compresses the higherfrequency
coefficients along the vertical axis, the
lower-frequency coefficients (those below about
1200 Hz) require interpolation to sufficiently inform
the brightness values for the multiple rows of a
single coefficient. In contrast, the higher-frequency
coefficients are under-sampled so that only
coefficients closest to display-rows are represented.
This optional nonlinear transformation of the
frequency axis allows the discrete time-frequency
spectrogram to be “warped” (different frequencies
stretched or compressed differently, but frequencyorder
preserved) without being “scrambled” (order
of represented frequencies not preserved)\cite{Oppenheim1972},
making it more amenable to visual pattern
recognition techniques\cite{Pucik2014, Harma2001, Jin2012, Giannoulis2013}.

\begin{figure*}
\includegraphics[scale=1.85]{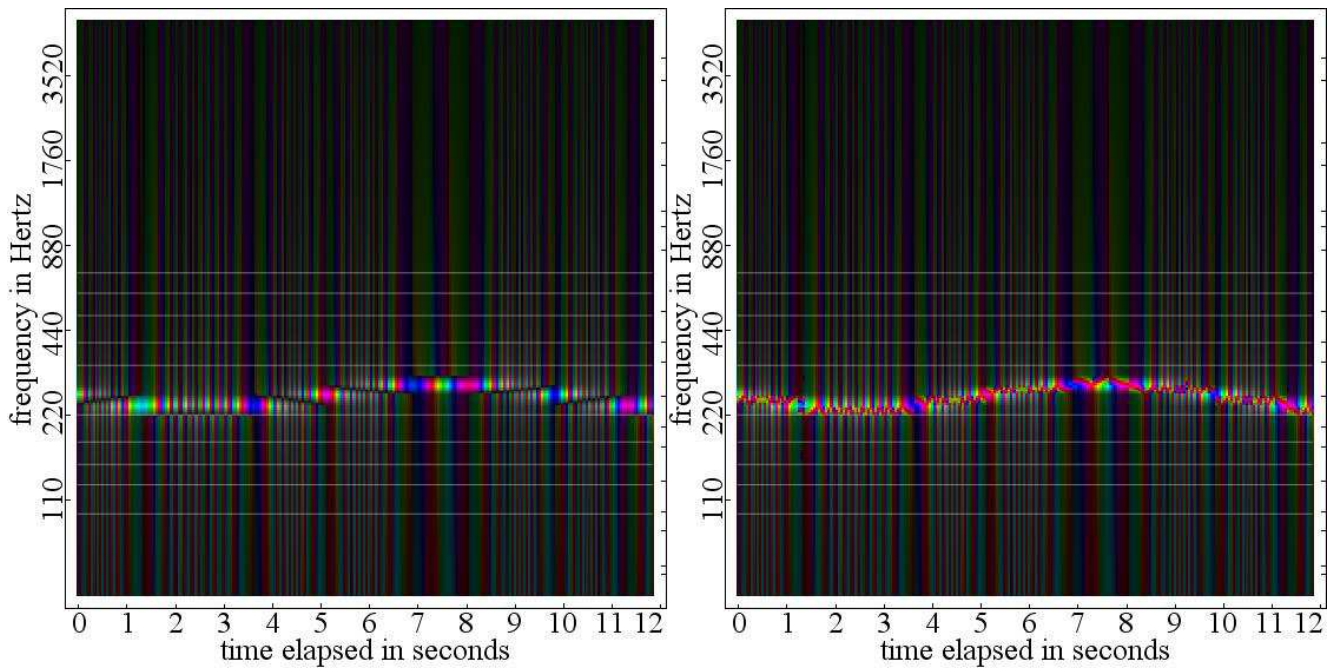}%
\caption{Rectangular (left) and polar (right) complex-color log-frequency interpolation of Fourier coefficients for a 10 percent frequency-modulated tone centered around 256 Hz.}
\label{fig2}
\end{figure*}

The log-frequency display is then rendered by first
completing the linear-frequency counterpart as
described above and then by mapping the vertical
axis to a log-frequency scale. At lower frequencies,
this requires interpolation between complex-valued
coefficients, for which there are two obvious
methods. While both polar and rectangular
interpolation routines were applied to this task,
rectangular interpolation (Fig. \ref{fig2}a) was found to be
preferable to polar interpolation (Fig. \ref{fig2}b). This is
because the rectangular approach produces a plot
that can be interpreted based on existing knowledge
of phases and coefficient centers, whereas the polar approach contains an inherent ambiguity in phase
assignment. The newly interpolated phase-angles are
then represented as colors as shown in Fig. \ref{fig1}.

Since each Fourier coefficient corresponds to a
frequency range determined by the FFT size, a
coefficient “center” is where a linear coefficient
index plots on the log-frequency scale. Since tiny
changes in amplitude can be detected by examining
more-sensitive phase-variations, mapping Fourier
phase to hue allows frequency-variations well below
the resolution allowed by a typical FFT size to be
visualized from one time-slice to the next as colored
stripes. In this way, rougher frequency data are
shown with brightness/saturation, while the finer
details are represented in color. Assuming a
sampling rate of 44.1 kHz and a 2048 FFT size, the
separation of coefficient centers is 44100/2048 ~
21.533 Hz.

At various points between coefficient centers,
rectangular interpolation results in zero-amplitude
phase-inversions. During these transitions, the
interpolated phases switch from being above the
center of the lower coefficient to being below the
center of the higher coefficient, or vice versa, at
which point the Fourier phase undergoes an
inversion. At these intersections, the interpolated
amplitudes reach zero before immediately becoming
positive again. The effect is that black lines appear
between coefficient centers with alternating color
rotations on either side. Such black lines are artifacts
of the rectangular phase-interpolation routine, and,
as an exception, do not actually correspond to zerointensities
in the input signal. This effect can be seen
in practice in Fig. \ref{fig2}a.

\section{Results}
\label{sec:result}

Realizations of this log-color visualization method
in HTML5/JavaScript\cite{Fraundorf2016w} have been shown to
process and render audio signals on a variety of
hardware platforms in $\simeq \frac13$ the time necessary to
maintain real-time synchronization. Since this
method for showing variation in phase among
Fourier coefficients allows for the representation of
a complex number by a single pixel, the entire FFT
can be conveniently displayed as a vertical line of
colored pixels with the brightness corresponding to
the log of the intensity of the Fourier coefficient and
the hue corresponding to the coefficient-phase. In
the time direction, steady variations in Fouriercoefficient
phase at the onset of each time-slice are
seen as colored stripes, with stripes of opposing
sequence (RGB vs. RBG) occupying opposite sides
of the zero-amplitude lines. When the oscillation
frequency is below the center of a coefficient, the
hue alternates in the RBG direction, and when the
oscillation frequency is above a Fourier-coefficient
center, the hue alternates in the RGB direction, as
seen in Fig. \ref{fig2}a. For a static tone, the frequency misalignment,
in Hertz, with the Fourier-coefficient
hardware-reference-frequency was found to be equal
to the number of color-cycles in a one-second
interval.

\begin{figure}
\includegraphics[scale=1.85]{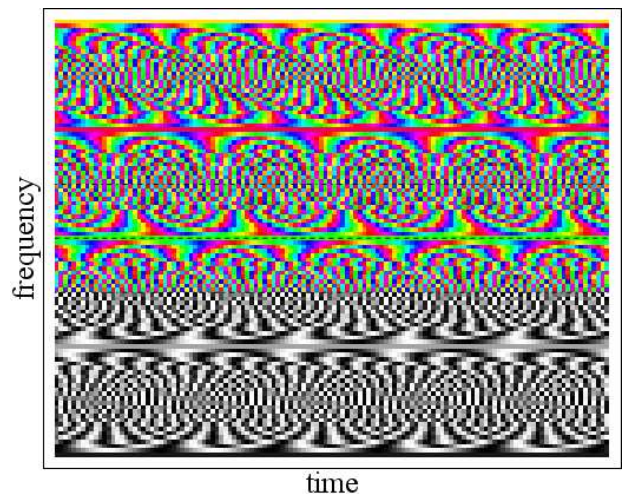}%
\caption{Composite beat-schematic, with 128
vertical time-slices arrayed across the horizontal
axis, and 4 center-to-center frequency-coefficients
on the vertical axis.}
\label{fig3}
\end{figure}

\begin{figure*}
\includegraphics[scale=1.9]{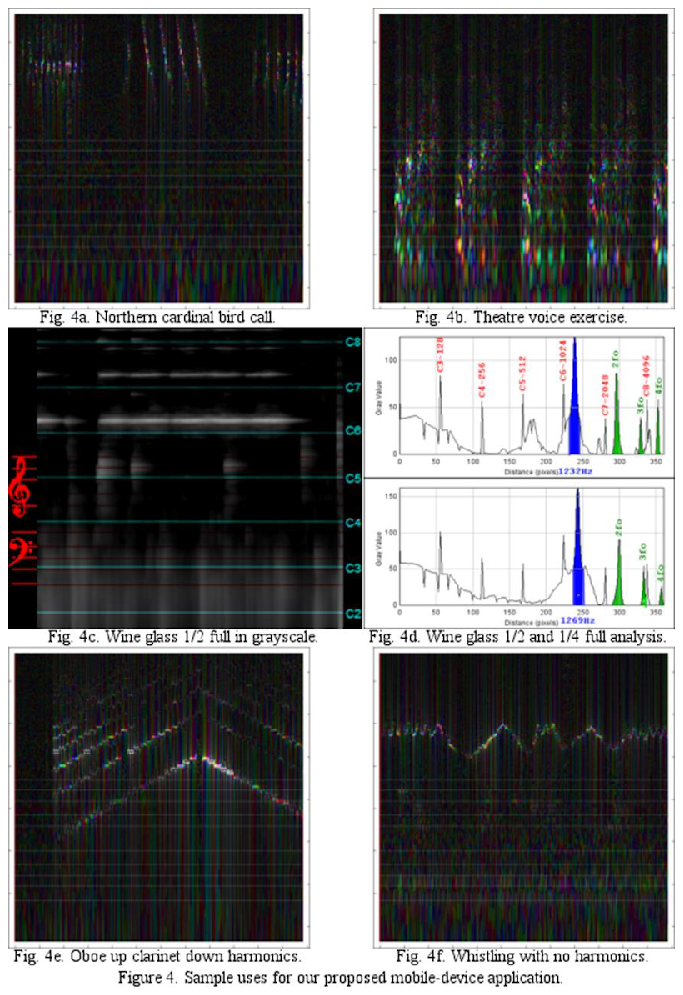}%
\caption{Sample uses for our proposed mobile-device application.}
\label{fig4}
\end{figure*}

Whenever the phase is centered on the Fourier
coefficient, the hue remains constant, which allows
highly accurate, well-centered data points to be
easily distinguished and isolated even in real-time.
In fact, the color-oscillations have a period inversely
proportional to the frequency offset from the
coefficient center, just as do amplitude beats used to
tune woodwind instruments (see Fig. \ref{fig3}). Each
frequency-coefficient in Fig. \ref{fig3} is divided into 25
lines with randomized phase-offsets to highlight beat-oscillations as a function of the frequencyoffset
from the coefficient-center (solid color lines).
The central dashed line in Fig. \ref{fig3} marks the center of
one frequency coefficient, with top and bottom
boundaries 1/8th of the height away in each
direction. The top 5/8ths of the plot show color
phase-beats with respect to coefficient center, while
the bottom 3/8ths shows monochrome amplitudebeats
with respect to a coefficient-centered note.

\section{Discussion}
\label{sec:discuss}

The connection of technologies like microphones,
digital displays, and computing power with
currently-existing, globally-interconnected, wireless
networks of highly-portable devices provides a
historically unique opportunity to drastically expand
the scope of applications for visual audio analysis. In
addition, versatile phase-sensitive audio-analysis
applications incorporating both modern (logfrequency)
and traditional (Grand Staff)
optimizations for enhancing visual pattern
recognition can provide a meaningful (or at least
relatable) basis from which anyone with experience
reading music can make interpretations of phasedetailed
audio data.

Several examples of applications involving these
features are illustrated in Fig. \ref{fig4}. In Figs. \ref{fig4}a-f the
vertical axis is frequency and the horizontal axis is
time. 

The inclusion of relevant sound images in text- or
print-based media (such as bird-sound field-guides
as suggested by panel in Fig. \ref{fig4}a) would allow users
without appropriate hardware to take advantage of
this technology by applying independent patternrecognition
analysis to existing sound-images.
Moreover such printed images may be used in
conjunction with, for example, a mobile-friendly
analysis-app to visually compare and classify live
captures with sound-visuals of known origin.

A real-time picture of incoming-sound (as in the
theatre voice example of Fig. \ref{fig4}b) can empower
voice imitators as well, even those who are hearingimpaired.
Figs. \ref{fig4}c and \ref{fig4}d illustrate the utility for
home experimenters in the spirit of Google's Science
Journal app, while Figs. \ref{fig4}e and \ref{fig4}f illustrate visual
comparison of musical instrument harmonics.

\begin{figure}
\includegraphics[scale=1.85]{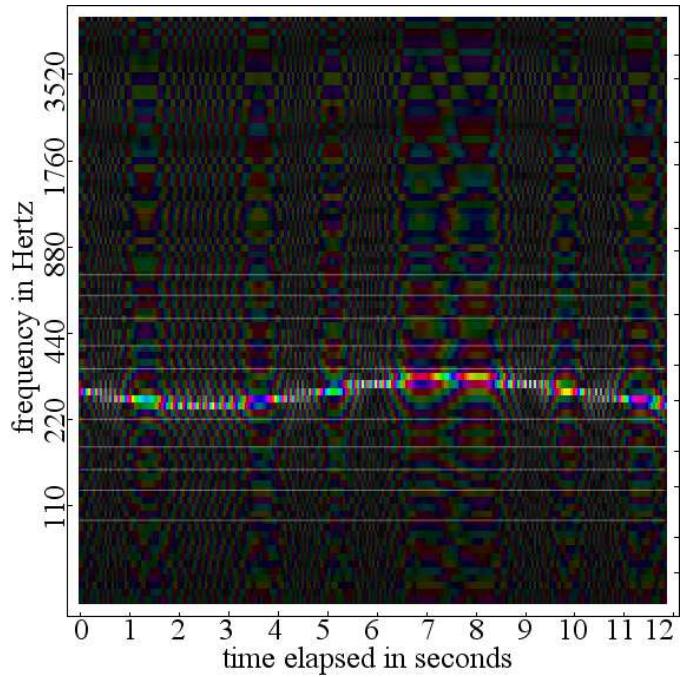}%
\caption{Half-note log-frequency rendition of a
10\% frequency-modulated tone centered around 256
Hz.}
\label{fig5}
\end{figure}

In addition to displaying data on the complete sound
wave, a generated image can be reverse-processed to
recover the original signal, including the original
phases imparted by the interference of the digital
detector with the source wave, which contain
information like relative angle to direction of sourcewave
propagation, etc. While CQTs have also been
shown to be invertible\cite{Velasco2011}, they do not display
phase information explicitly and generally require
additional computational resources compared to the
discrete FFT\cite{Schorkhuber2010}. Since musical notation provides a
practical reference, and since each pixel can be
mapped back to the original sound, both human
imitation and recovery to audio are also possible.
Other modifications, such as adjustment of the
frequency axis so Fourier coefficients match
frequencies of particular tuning standards, could be
used to readily display whether a note is in
appropriate tune, or if not, whether it is sharp or flat
and by precisely how much. Such note-specific
applications would be completely accessible to
anyone who reads music, and would incorporate a new class of potential users of technically
sophisticated audio analysis software.

Finally, our browser implementations are only one
facet of this development. More specialized
implementations, e.g., in hardware instead of
software will open the door to other uses. For
instance, by doing a separate transform for each
half-note in a log-frequency display, one can avoid
all interpolation artifacts and put any sound into
playable music notation (see Fig. \ref{fig5}). In fact, a single
12 second multi-octave chromatic scale could be
used to quantify the tuning state of all notes on a
piano.

\section{Summary}
\label{sec:summary}

By enabling sound visualization that includes
source-detector phase-interference in a convenient,
familiar, and portable format, this combination of
processing and display techniques opens the door for
improved accuracy in sound measurement and
analysis in a plethora of new and diverse
environments and applications. With further
development of robust audio visualization software
in parallel with semiconductor technology, the
general public will soon have access to a growing
variety of specialized, phase-interferometric tools to
record, analyze, and recreate sounds on an
increasingly real-time basis. As software continues
to be developed, applications which take advantage
of traditional musical notation will always have the
advantage of wider accessibility by the general
public, as well as additional potential for musical
reproduction and conceptual reference. Furthermore,
the ability to record and analyze audio in a visual
form that retains precise information regarding the
physical orientation of the actual sound wave in
space relative to the detector that recorded it marks
an important step in detailed sound-feature analysis.


\begin{acknowledgments}
Thanks to the regional nanoscience community for
diverse applications of "spatial-periodicity"
complex-color, which resulted in the basic idea here
by PF, plus programming and paper-writing by SW.
\end{acknowledgments}


\bibliography{ifzx3}

\end{document}